\documentclass[preprint,pra,aps,amsmath,amssymb,showpacs,superscriptaddress]{revtex4-1}
\usepackage{upgreek}
\usepackage{amsfonts}
\usepackage{amssymb}
\usepackage{amsmath}
\usepackage{graphicx}
\usepackage{soul,xcolor}
\usepackage{dcolumn}
\usepackage{bm}
\usepackage[T1]{fontenc}
\usepackage{hyperref}
\usepackage[normalem]{ulem}
\usepackage{gensymb}

\begin{document}

\title{Graphene resonator as an ultrasound detector for generalized Love waves in a polymer film with two level states }

\author{Antti Laitinen}
\affiliation{Low Temperature Laboratory, Department of Applied Physics, Aalto University, Espoo, Finland}
\author{Jukka-Pekka Kaikkonen}
\affiliation{Low Temperature Laboratory, Department of Applied Physics, Aalto University, Espoo, Finland}
\author{Thanniyil S. Abhilash}
\affiliation{Low Temperature Laboratory, Department of Applied Physics, Aalto University, Espoo, Finland}
\author{Igor Todoshchenko}
\affiliation{Low Temperature Laboratory, Department of Applied Physics, Aalto University, Espoo, Finland}
\author{Juuso Manninen}
\affiliation{Low Temperature Laboratory, Department of Applied Physics, Aalto University, Espoo, Finland}
\author{Vladislav Zavyalov}
\affiliation{Low Temperature Laboratory, Department of Applied Physics, Aalto University, Espoo, Finland}
\author{Alexander Savin}
\affiliation{Low Temperature Laboratory, Department of Applied Physics, Aalto University, Espoo, Finland}
\author{Andreas Isacsson}
\affiliation{Department of Physics, Chalmers University of Technology, SE-41296, G\"{o}teborg, Sweden}
\author{Pertti J. Hakonen}\email[Corresponding author: pertti.hakonen@aalto.fi]{}
\affiliation{Low Temperature Laboratory, Department of Applied Physics, Aalto University, Espoo, Finland}

\begin{abstract}

We have investigated surface shear waves at 22~MHz in a
0.5-micron-thick polymer film on SiO$_2$/Si substrate at low
temperatures using suspended and non-suspended graphene as
detectors. By tracking ultrasound modes detected by oscillations of a
trilayer graphene membrane both in vacuum and in helium superfluid, we
assign the resonances to surface shear modes, generalized Love waves,
in the resist/silicon-substrate system loaded with gold. The
propagation velocity of these shear modes displays a logarithmic
temperature dependence below 1~K, which is characteristic for
modification of the elastic properties of a disordered solid owing to
a large density of two level state (TLS) systems. For the dissipation
of the shear mode, we find a striking logarithmic temperature
dependence, which indicates a basic relation between the speed of the
surface wave propagation and the mode dissipation.

\end{abstract}

\pacs{}

\maketitle

\section{Introduction}
Graphene forms a promising platform for a multitude of detectors and
sensors~\cite{Ferrari2015}. One classical application is strain gauges
which can be formed using graphene in a polymer
matrix~\cite{Boland2016}. Suspended graphene can also be employed as a
strain gauge for deformations, while it simultaneously allows one to
gauge the environment via chemical/physical attachment, pressure, and
viscous damping. These properties make suspended graphene devices
versatile tools at low temperatures~\cite{Hone2013}. This includes
detectors of ultrasound~\cite{Verbiest2018} with sensitivity of the order of $10^{-12}\, \text{m}/\sqrt{\text{Hz}}$, and ultrasonic surface
waves in this work utilized as a probe of two level systems. Surface wave devices are of
considerable potential for sensing applications in liquid
environments. Interaction of a liquid or solvent with the sensor
surface can modify the properties of acoustic waves (velocity,
frequency, or phase) which facilitates detection of changes in the
liquid properties, or alternatively, sensing of adsorbed
molecules. Rayleigh wave devices display superior sensitivity in
general reaching sensitivities of the order of $10^{-17}\, \text{m}/\sqrt{\text{Hz}}$ using interdigitated capacitors and $10^{-15}\, \text{m}/\sqrt{\text{Hz}}$ using optical interferometric techniques in the MHz frequency range \cite{Boltz1996}. However, they appear unfit for operation in liquid due to strong
damping associated with radiation losses into the
liquid~\cite{Harding1997}. Hence, alternative measurement schemes are of interest.

Recently, sensing devices based on horizontal shear waves (SH)
confined to the surface, Love waves, have been developed (see
e.g. Refs.~\onlinecite{Kovacs1992a,Harding1997,Rocha2013}, and
references therein). Love waves can propagate near the surface of a
substrate, when the surface is covered by a thin film having suitable
properties for guiding the waves. The primary condition for existence
of Love waves is that the shear wave velocity in the film is smaller
than the shear wave velocity in the substrate.  This can be achieved,
e.g., by using polymer resists like PMMA (poly methyl
methacrylate)~\cite{Gizeli1992a,Harding1997} or SU-8 (epoxy-based
polymer)~\cite{Roach2007} which behave as soft viscoelastic materials
with a quite low shear resistance, and hence their acoustic shear wave
velocities are small, $v_{\rm sh} \simeq 1100\,$m/s for PMMA at room
temperature~\cite{Gizeli1992}. Small density ($\rho \sim
1\,$g/cm$^{3}$) and small $v_{\rm sh}$ lead to good isolation of the
polymer surface shear waves from the substrate.  The only extrinsic
loss mechanism for a shear wave mode interacting with a liquid is a
result of viscous surface coupling, i.e. no mode conversion is
involved, contrary to the case of Rayleigh
waves~\cite{Auld1973}. Since surface shear waves do not have elastic
coupling into liquids, these waves are particularly attractive for
sensing systems in which the liquid environment only acts as the
transporting medium for the searched compounds.

In our work, we are interested in lift-off-resist polymer films (made
of LOR~3A polydimethylglutarimide) which are quite commonly in use in
graphene nanofabrication. These LOR films are employed as support
layers for suspended graphene devices reaching ultimate carrier
mobility. In this work, we have operated our suspended graphene
devices as mechanical resonators, both under vacuum conditions as well
as immersed in $^4$He superfluid. By comparing the results of devices in
vacuum and immersed in superfluid, we can distinguish between
graphene, gold, and substrate modes. We find that an approximately
500~nm thick LOR polymer film on top of a regular Si/SiO$_2$ substrate
display resonances corresponding to SH-waves confined to the surface,
which can be considered as generalized Love
waves~\cite{McHale2002}. The Love wave modes distinguish themselves from the other observed modes by not displaying a frequency shift when immersed in superfluid helium, as the effective mass of shear modes remains unchanged. We have investigated these waves as a
function of temperature and find good agreement between the observed
properties and the predictions from two level state (TLS) tunneling
models \cite{Auld1973,Roach2007}. Owing to the high purity of
the silicon/SiO$_2$ substrate, we conclude that the TLS systems have
to reside in the topmost, amorphous resist layer.

\section{Experimental methods}
The layout of our samples is quite standard for suspended graphene
devices, consisting of bonding pads, metallic leads that are partly
suspended, and a fully suspended graphene section.  However, in this
work the bonding pads of $150 \times 150$ $\upmu$m$^2$ area are
considered as surface wave generators.

Owing to the back-gated structure, the capacitive
  force becomes large across the LOR and SiO$_2$ layers, which
  facilitates excitation of resonance-enhanced deformations of the
  substrate. For the Love-waves we are concerned with here, we seek to
  excite resonances that give rise to SH-modes localized near the
  surface.  The simplest source of such SH-resonances is the pad
  itself executing resonant horizontal vibrations. This problem,
  vibrations of rectangular foundations on stratified elastic and
  viscoelastic media, has a long history in the context of seismic
  safety engineering~\cite{Kobori1971, Luco1974}. Although it is
  possible to solve the problem numerically~\cite{Wong1976,Ghadi2013},
  we note that the characteristic frequency can be estimated via the
  static compliance for shear
  deformations~\cite{Kobori1971,Gazetas1991}. We obtain in this way
  the estimate:
    \begin{equation}
    f_{\rm res}\approx (2\pi)^{-1}\sqrt{\frac{9Ga}{(2-\nu)M_{\rm mode}}}\approx 24\,{\rm MHz}.
    \end{equation}
  Here $2a=150\,\upmu$m is the sidelength of the Au/Cr pad, $M_{\rm
    mode}\approx 33\,$ng is the effective mass of the resonating mode, $G=1.9\,$GPa the shear
  modulus of the LOR, and $\nu=0.35$ its Poisson's ratio.
  Importantly, this resonant frequency is proportional to the velocity
  of SH-waves $v_{\rm sh}=\sqrt{G/\rho}$ in the substrate, which for
  LOR becomes $v_{\rm sh}\approx 1400\,$m/s.  With Young's modulus
  $E_{\rm LOR}=5\,$GPa ~\cite{Foulds2008} for the LOR layer and dielectric constant $\epsilon_r
  \simeq 3$, we find for a typical gate excitation ($V_{\rm DC}=5\,$V
  and $V_{\rm AC}=5\,$mV) that the resonant amplitude is in the
  picometer range in our geometry. COMSOL simulations further confirm
  that the lowest eigenfrequencies of the gold pads involve mostly
  lateral movement while vertical displacement in opposite directions
  is found foremost at opposing edges. The movement can well be
  detected using graphene resonators, which display non-linear
  response already at an oscillation amplitude $\gtrsim 50$ pm for
  flexural modes~\cite{Song2012}.

The distance from the wave generators to the sample, located on the
symmetry axis (diagonal to the square) of the pad, is 100$\,\upmu$m for
the two pads, respectively. The sample itself is oriented at
45$\degree$ angle with respect to the symmetry axis (see
Fig.~\ref{sample}b). For horizontal angular frequency excitations with
$\omega_{\rm sh}$ larger than the characteristic frequency scale
$v_{\rm sh}/a$, the surface wave field may undergo considerable
phase-shifts ($\gtrsim90\degree$) between pad edge and
detector~\cite{Ghadi2013}. Such a large phase shift can be considered as an
indication of spatially separated generation and detection of the
waves. The separation also means that additional frequency and
temperature-dependent damping may occur before detection.


\subsection{Samples and their characterization}
\begin{figure}[tbp]
\includegraphics[width=7.4cm]{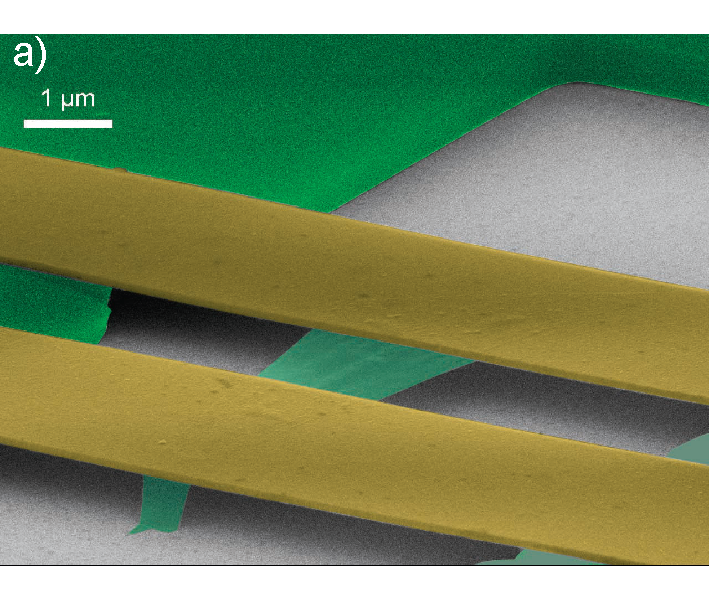}
\includegraphics[width=5.8cm]{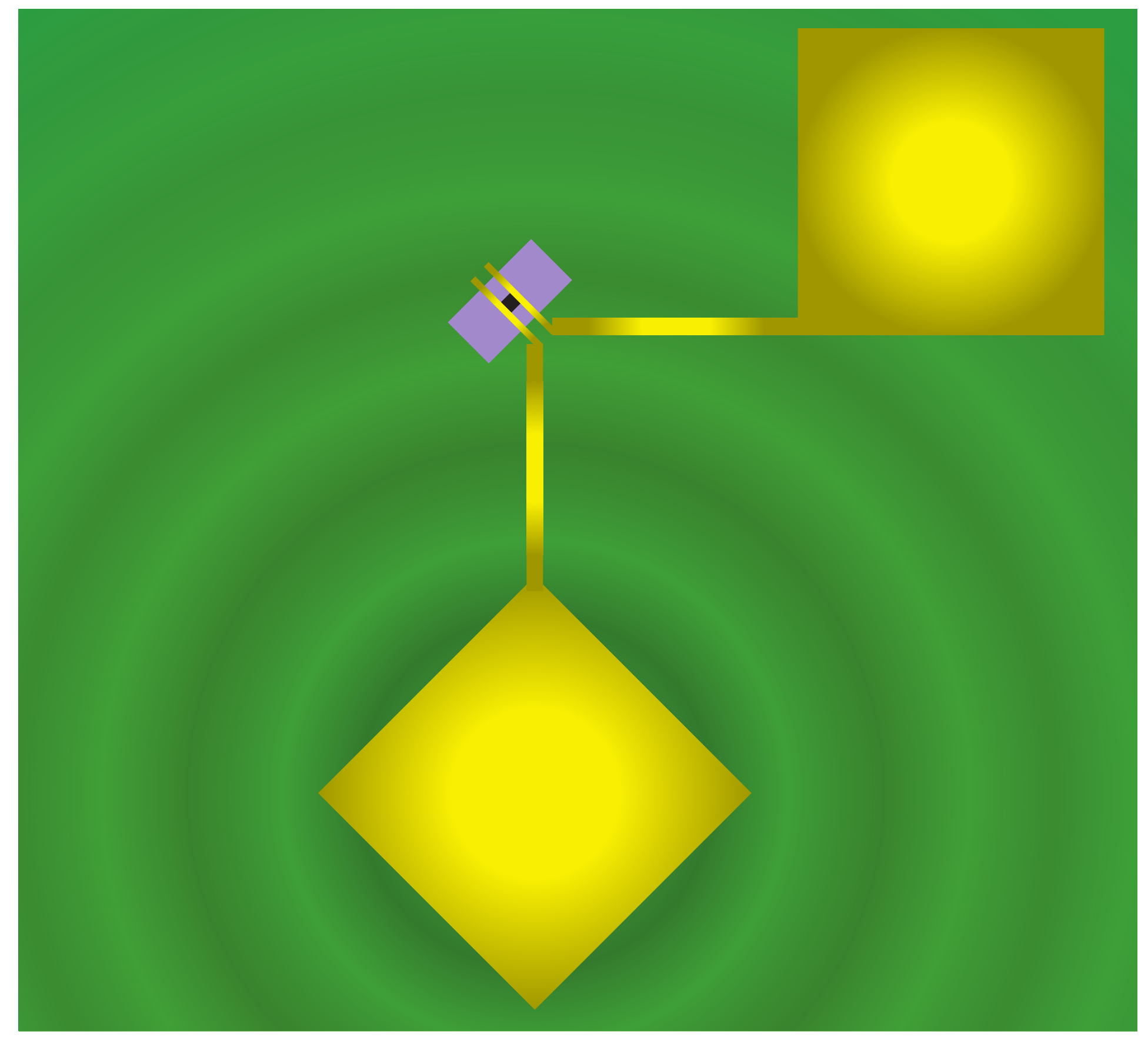}
\caption{(a) Scanning electron micrograph of the trilayer graphene
  resonator sample: dark green denotes LOR resist supports, yellow
  indicates two gold contacts that are suspended over a distance 3$\,\upmu$m in the
  center, while light green denotes the suspended graphene. (b)
  Schematic layout of our samples from the top: 500-nm-thick LOR
  resist (green) lays on top of 285-nm-thick SiO$_2$ layer (purple),
  and gold leads (yellow) on top of LOR. Surface waves are generated
  from one of the ponding pads.}
\label{sample}
\end{figure}

Our sample fabrication follows the scheme based on different
selectivities of resists: lift-off-resist (LOR) for support and PMMA for
lithography were employed in the sample fabrication (for
details see Refs.~\onlinecite{Kumar2018}
and~\onlinecite{Tombros2011}). The total thickness of the LOR layer
amounted to $\sim 500\,$nm after two spin-coatings. Then, we exfoliated
graphene (Graphenium, NGS Naturgraphit GmbH) using a heat-assisted
technique~\cite{Huang2015}. Graphene flakes were located on the basis
of their optical contrast, and their numbers of atomic layers were verified
using a Raman spectrometer with He-Ne laser (633~nm). The top metallic
contacts (Cr/Au, 5$\,$nm/60$\,$nm) defining the freely suspended area of
graphene (see Fig.~\ref{sample}a) were patterned and deposited using
e-beam lithography and ultra-high vacuum metal evaporation
techniques. Then, the LOR under and around the graphene flake was
removed by exposing it to a 20~kV e-beam and developing in ethyl
lactate. Finally, the chip was lifted off from hexane bath where
hexane's low surface tension allowed the graphene membrane to pass the
liquid surface unharmed. A strongly doped silicon Si++ substrate
(thickness $525\,\upmu$m) with a 285~nm layer of thermally grown
SiO$_2$ provided the back gating electrode for the sample. The
geometry of the sample is illustrated in Fig.~\ref{sample}.

\begin{figure}[tbp]
\includegraphics[width=7.7cm]{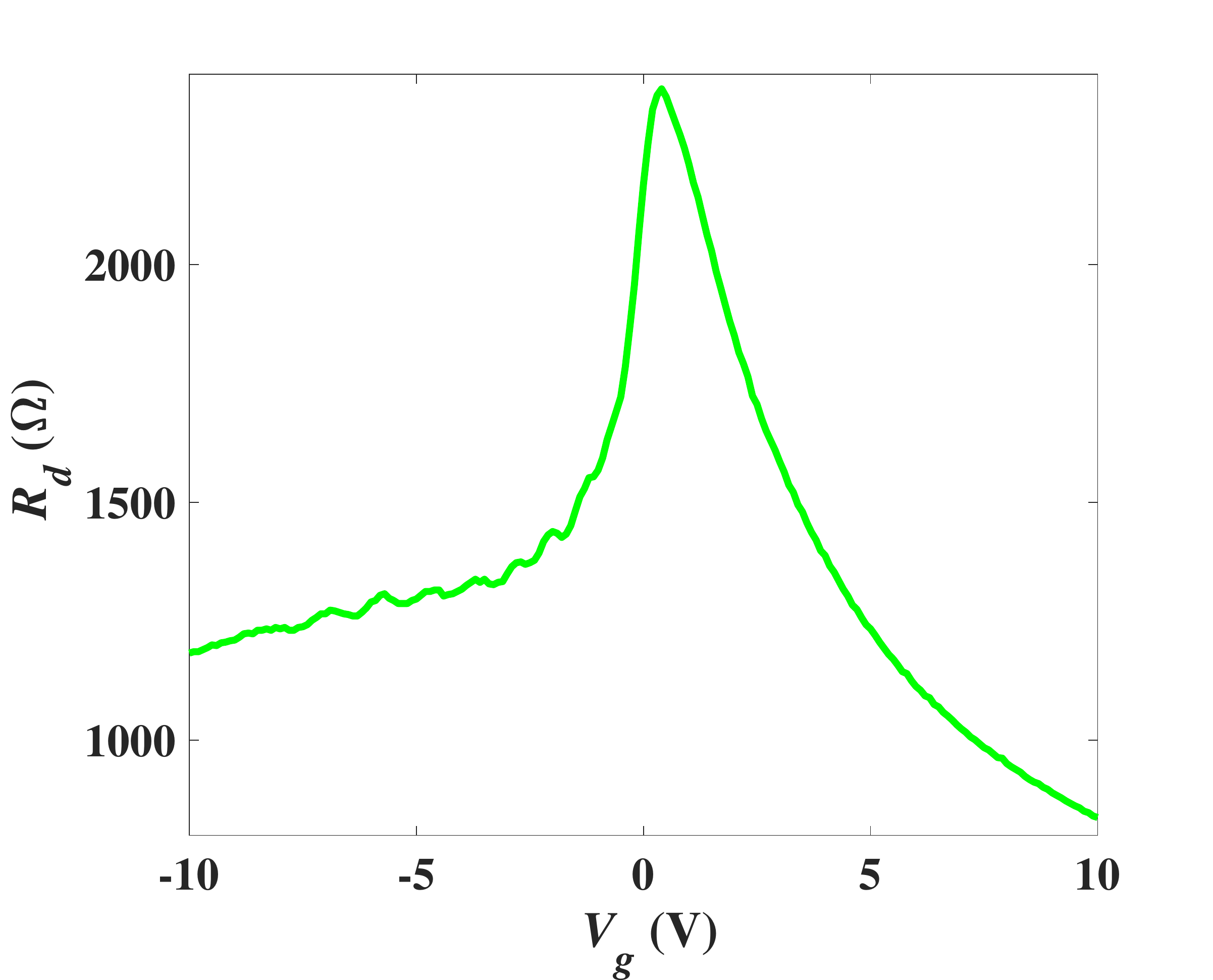}
\caption{Zero bias resistance vs. gate voltage $V_{\rm g}$ for $1.7 \times
  1.8\,\upmu\text{m}^2$ trilayer sample displayed in
  Fig.~\ref{sample}a. The Dirac point is located at $V_{\rm g}= +0.4\,$V. }
\label{fig:GateSweep}
\end{figure}

Initially, after the fabrication process, our suspended devices tend
to be $p$-doped in the first resistance $R$ \textit{vs.} gate voltage
$V_{\rm g}$ scans. Following the initial characterization, the samples
were cooled down to $T$ = $10\,$mK base temperature of a Bluefors
LD-400 dry dilution refrigerator. Prior to electrical
characterization, all devices were current annealed at the base
temperature. These samples on LOR were typically annealed at a bias
voltage of 1.6$\pm$0.1$\,$V which is comparable with the optimal
annealing voltage of our HF etched, rectangular two-lead
samples~\cite{Laitinen2014} and are close to the values employed for
our monolayer samples on LOR~\cite{Kumar2018}. The trilayer graphene
device here is effectively slightly $p$-doped even after
annealing. Nevertheless, $pn$-junctions are formed on the negative
gate voltage side, which is visible as larger resistance at $V_{\rm g}
= -10\,$V than at $V_{\rm g}=+10\,$V, see
Fig.~\ref{fig:GateSweep}. Besides a small difference in the work
functions of the materials~\cite{Laitinen2016}, we assign this
behavior to non-uniform doping and screening across the graphene
layers.

Standard lock-in techniques using DL Instruments 1211 preamplifier
(with sensitivity $10^{-5}\,$V/A) followed by a Stanford SR830 lock-in
were employed for conductance measurements. The sample was voltage
biased with $V_{\rm b} =8\,\upmu$V source-drain AC voltage while the
back gate was tuned from $V_{\rm g} = -10\,$V to $V_{\rm g} =
+10\,$V. Electronic mobility in the devices could be determined from
the conductance measurements $G(V_{\rm g})$, and was found to be $\mu
\sim 10^4\,$cm$^2$/Vs for our trilayer sample. The gate voltage was
converted into charge carrier density using $n = (V_{\rm g}-V_{\rm
  g}^{\rm D})C_{\rm g}/e$, where $V_{\rm g}^{\rm D}$ denotes the
offset of the Dirac point from $V_{\rm g} = 0\,$V. The investigated
trilayer sample had $V_{\rm g}^{\rm D} = +0.4\,$V (see
Fig.~\ref{fig:GateSweep}). For our suspended monolayer graphene
samples that were fabricated and characterized in the same manner as
the trilayer sample, we found the Dirac point close to $V_{\rm g}^{\rm
  D} \approx -1\,$V, and mobility exceeding $10^5\,$cm$^2$/Vs. We also
fabricated and measured a LOR-supported control sample where the
fabrication and characterization was identical to the suspended samples
except for omitting the final LOR removal step. This sample
had a mobility around $2000\,$cm$^2$/Vs and the Dirac point $V_{\rm g}^{\rm D} >
+100\,$V.
\subsection{Detection of mechanical modes}
For the detection of mechanical modes, we employed the mixing current
method based on frequency
modulation~\cite{Gouttenoire2010}. Typically, as considered in
Ref.~\onlinecite{Gouttenoire2010}, the mechanical motion yields a
mixing current via the change in differential conductance $G_{\rm d}(f)$ due to
charge modulation $C_{\rm g} V_{\rm g} \delta z/z$ induced by an amplitude change $\delta
z$ as:
\begin{eqnarray} \label{fitfunc}
  I_{\text{mix}} \propto \frac{{{\partial}G_{\rm d}(f)}}{{\partial z}}\left| {\frac{{\partial {\mathop{\rm Re}\nolimits} (z)}}{{\partial f}}} \right|, \\
 \nonumber \delta z = z_0 \frac{\exp(-i \varphi)}{f_{\text{res}}^2 - f^2 + if_{\text{res}} Q f}.
\end{eqnarray}
Here $\delta z$ describes the resonant response of the resonator, $z$
denotes the distance between the graphene membrane and the gate
electrode.  The sample is biased using a frequency modulated signal
$V^{\rm FM}(t)=V^{\rm AC} \cos(2\pi ft+(f_{\rm A}/f_{\rm L})\sin(2 \pi
f_{\rm L} t))$; $V^{\rm AC}$ is the amplitude of the carrier frequency
signal, $f_{\rm L}$ is the frequency modulation (typically 613~Hz was
employed), and $f_{\rm A}$ defines the modulation depth (around
1-5~kHz). The angle $\varphi$ indicates a phase shift between the
drive and the response, which e.g. can be 0 or $180\degree$
for graphene oscillations driven from the clamps
  due to the slow speed of flexural modes \cite{Juuso}. Note that
this mixing current formula works also for mechanical displacements in
other directions provided that there is proper conductance modulation
by strain.

The appealing property of the FM modulation method is its
insensitivity against spurious resonance signals, when compared with
other detection schemes, for example the rectification
method. Thus, non-linearities in the IV curve of the graphene device
cannot give rise to the observed mechanical resonance spectra.

\section{Experimental Results}

\begin{figure}[tbp]
\includegraphics[width=7.9cm]{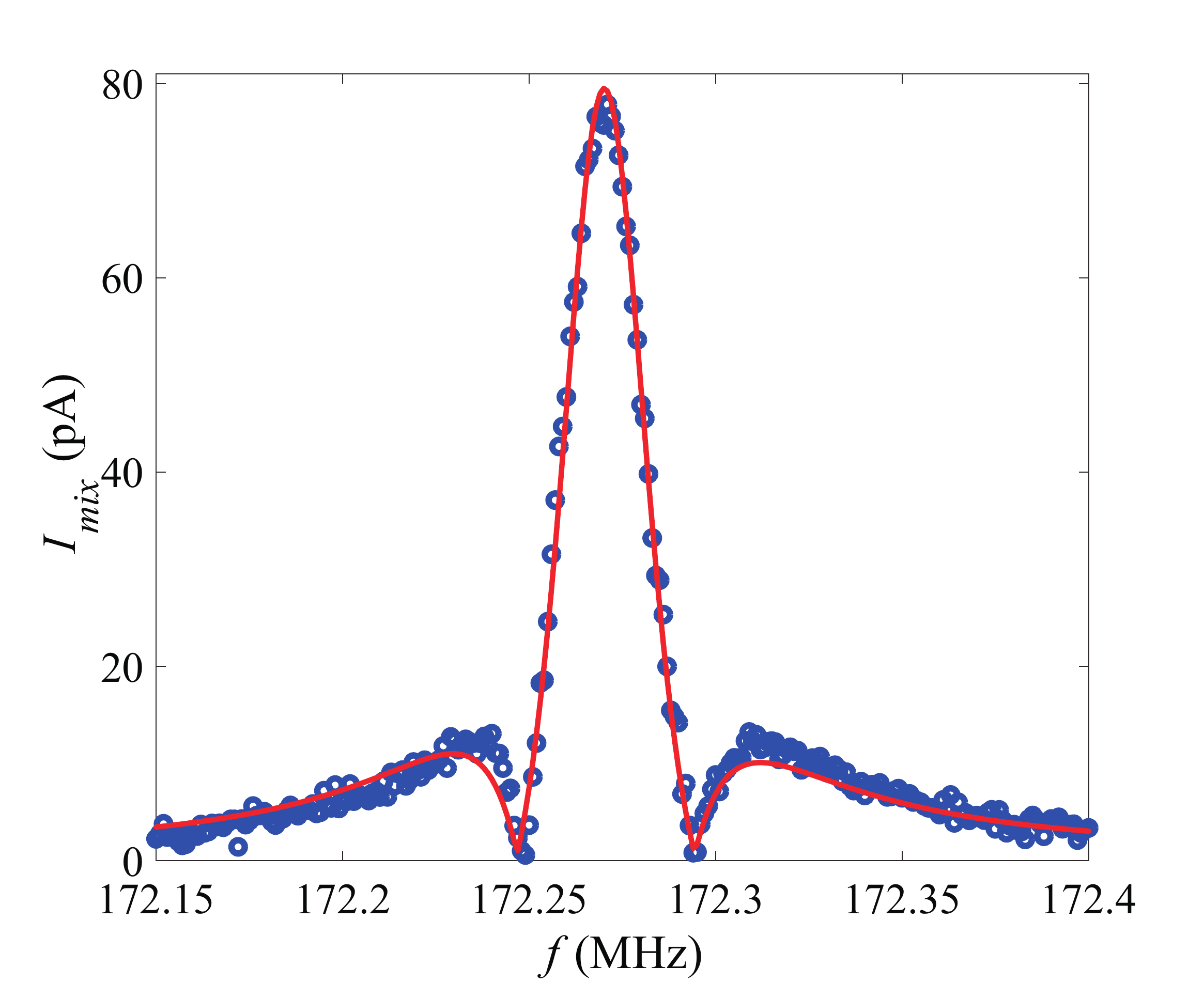}
\includegraphics[width=8.3cm]{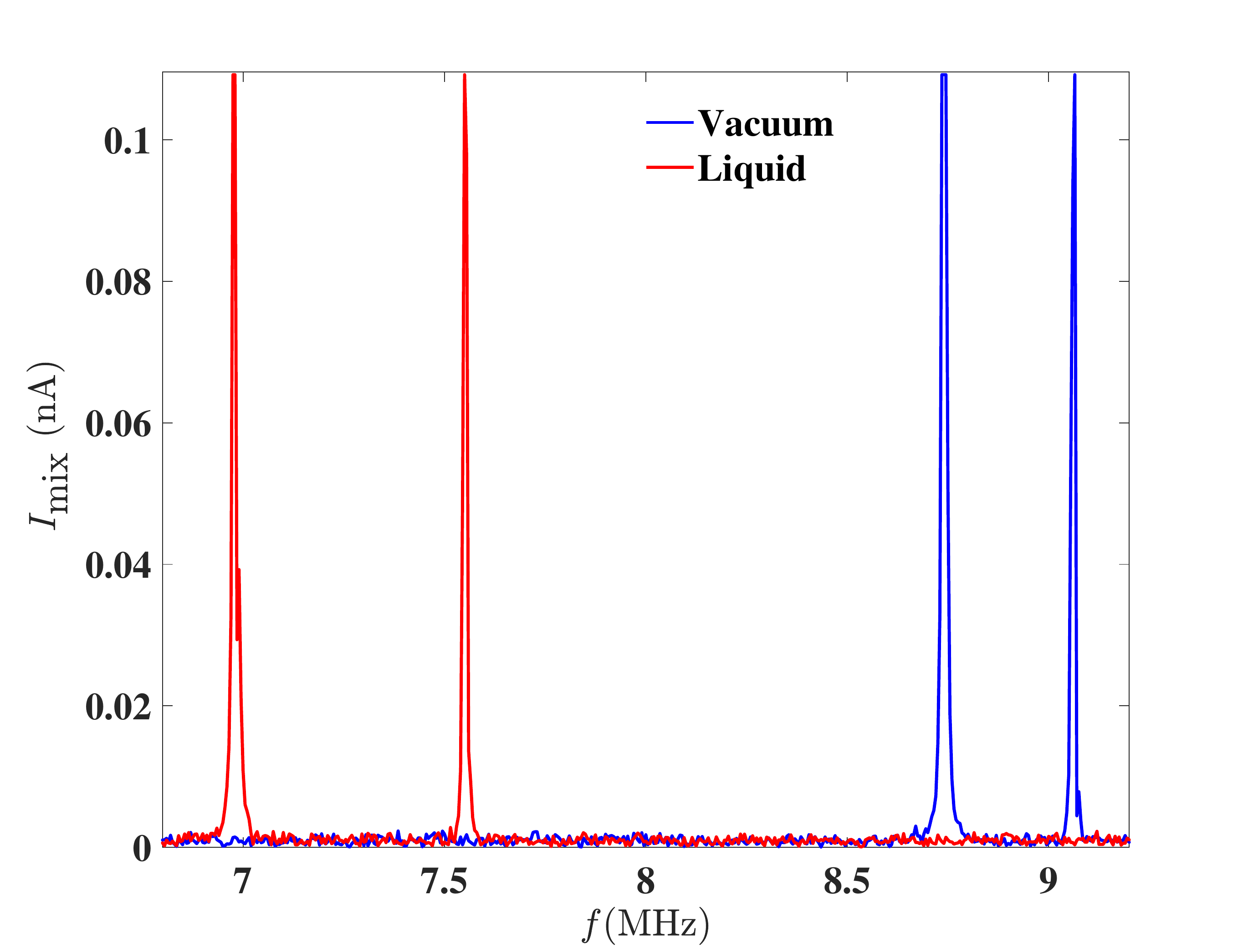}

\caption{(left) Mechanical resonance shape for the fundamental mode in
  trilayer graphene. The spectrum was measured at $T=4.5$ K using an
  excitation of 1 mV$_{rms}$. (right) Mechanical resonances of the
  gold beams, both in vacuum (blue) and in liquid (red). For details,
  see text.}
\label{fig:resoShape}
\end{figure}

We scanned the frequency range $3 - 500$~MHz for mechanical
resonances, both in vacuum as well as in superfluid
$^4$He. Altogether, we found about 30 resonances, the majority at
frequencies above 100~MHz. Owing to their high frequencies and gate
voltage dependencies, we assigned most of these modes to the
graphene. The resonance identified as the fundamental mode in our
trilayer graphene sample is illustrated in
Fig.~\ref{fig:resoShape}a. This spectrum has the regular symmetric
form of a FM modulated resonance signal. The overlaid red curve is
obtained from Eq.~\ref{fitfunc} using $\varphi=0$. The fit yields for
the quality factor $Q=3500$ ($T=4.5\,$K).

Fig.~\ref{fig:resoShape}b illustrates resonance peaks belonging to the
the gold leads. They appear via the motion of graphene and their shape
has a phase shift ($\varphi \neq 0$) compared with the graphene
resonances. We display only the resonances assigned to the flexural
modes of the gold at $f_{\text{res,Au}}=9.1$ and 8.7~MHz in
vacuum. Above 10~MHz, we observe also torsional modes of the gold
beams. The gold modes are coupled via the graphene sheet, which leads
to complicated internal dynamics of our system at large amplitudes.

When the graphene device is immersed in liquid, the mechanical modes
change dramatically. In particular, the frequency of the graphene
modes drops to about $10\,$\% of their original frequency and their
amplitude goes down by an order of magnitude. The change for both
graphene and gold resonance frequencies arises due to the enhanced
effective masses, which for plates with large aspect ratio is quite
substantial~\cite{Blaauwgeers2007}. The frequency of the gold beam
resonances decreases by $\sim 1.6\,$MHz, i.e. by approximately $16\,$\%,
which agrees well with estimates based on
Ref.~\onlinecite{Blaauwgeers2007}.
%
%

\begin{figure}[tbp]
\includegraphics[width=8cm]{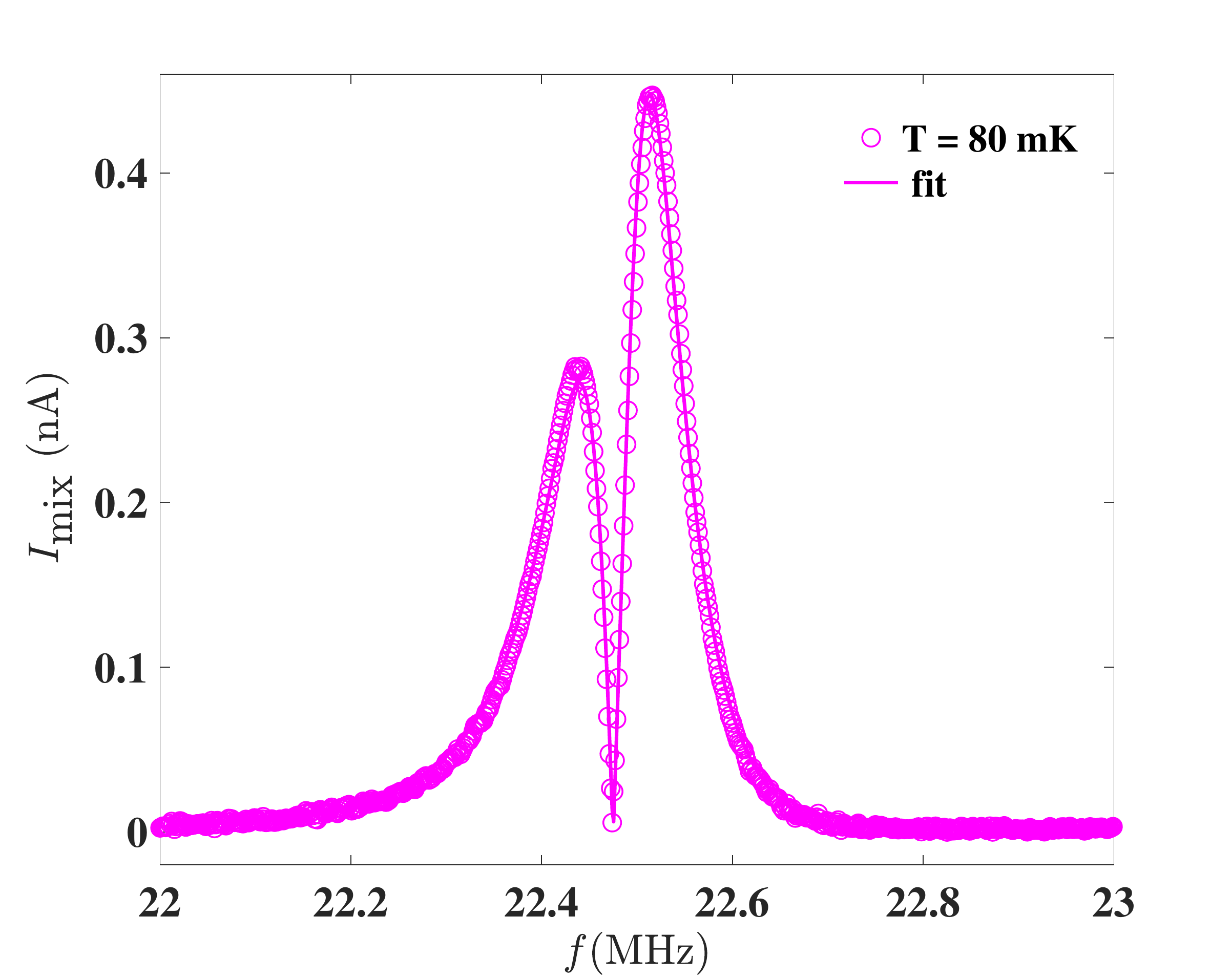}
\includegraphics[width=8cm]{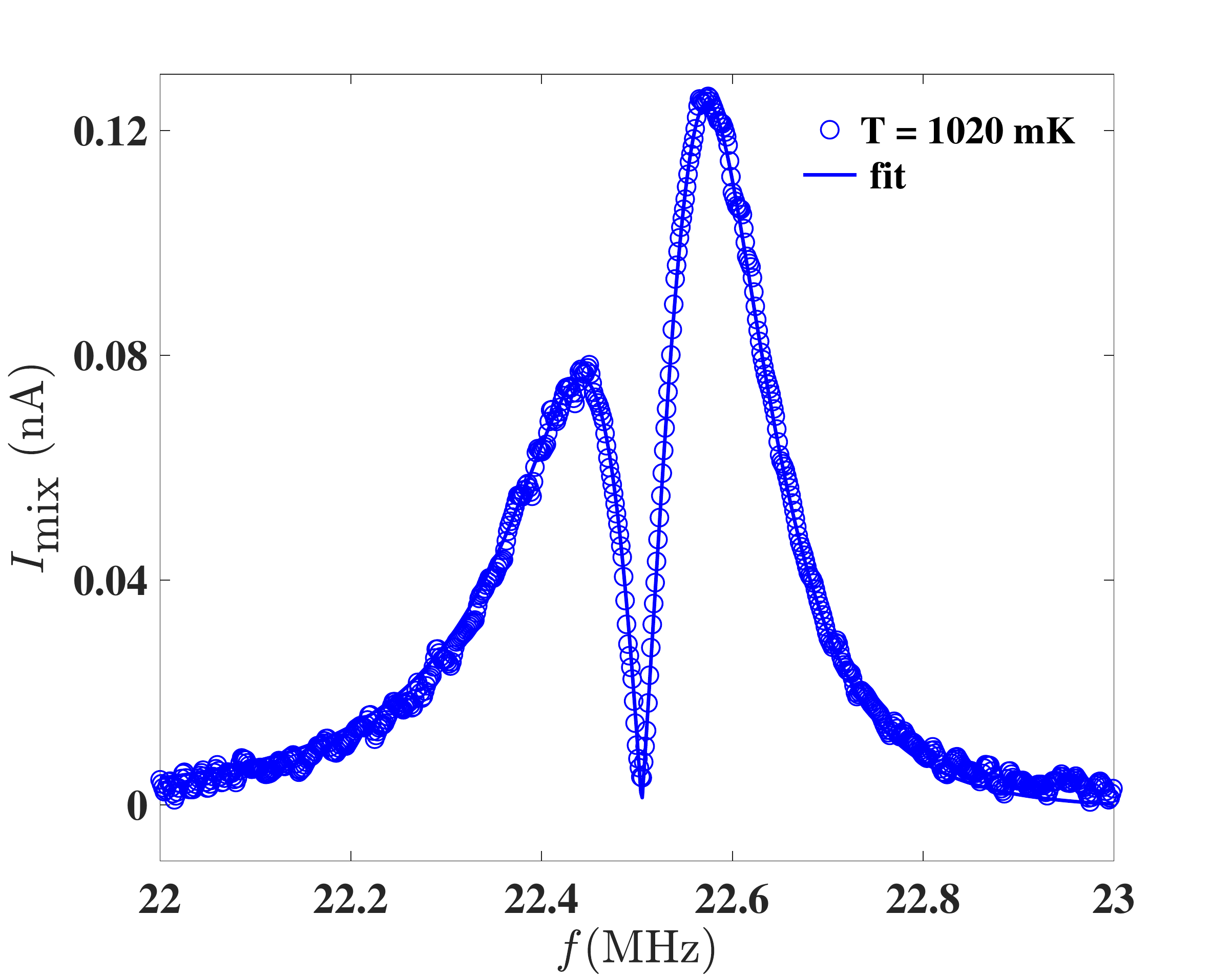}
\caption{Mechanical resonance shape for an asymmetric resonance at
  22.5~MHz, measured at $T=80\,$mK (a) and at $T=1020\,$mK (b). The
  sample chamber was filled with superfluid $^4$He at pressure $p =
  0.5\,$bar.}
\label{fig:resoShape2}
\end{figure}

Our data also include resonances that do not change at low
temperatures within our experimental accuracy of $2-3$~kHz, when the
experimental chamber is filled with superfluid helium. The spectrum of
one such mode around $f_{\text{res}} = 22.5$ MHz is displayed in
Fig.~\ref{fig:resoShape2}a at $T=80\,$mK: the shape is asymmetric
which can be accounted for by a phase shift between the drive and the
response. The solid curve is a fit to the resonance formula in
Ref.~\onlinecite{Gouttenoire2010} using $\varphi \sim -64\degree$ for
mixing the in-phase and out-of-phase responses. This phase shift is
well in line with propagating waves between the pad and the graphene
detector.  Fig.~\ref{fig:resoShape2}b illustrates the same resonance
at $T=1.02\,$K. There is a small increase in frequency, line width,
and in $\varphi$, which we have obtained using Eq.~\ref{fitfunc}. Also, the
amplitude of the resonance grows more strongly with lowering
temperature than the inverse width of the mode. This is a sign that
the detection sensitivity of the graphene membrane for this mode
increases when the dissipation of the generated sound decreases around
22.5~MHz.

\begin{figure}[tbp]
\includegraphics[width=8.1cm]{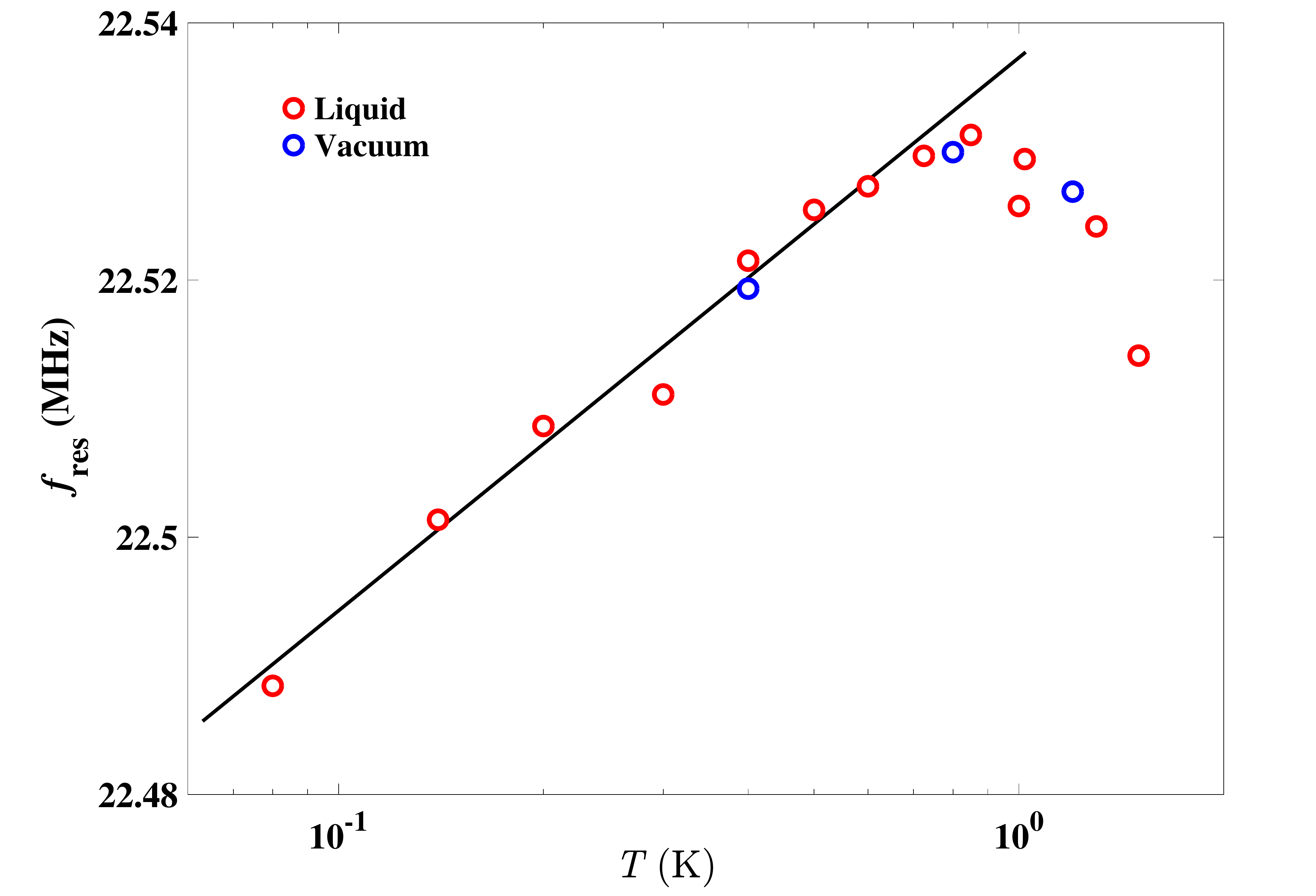}
\includegraphics[width=8.1cm]{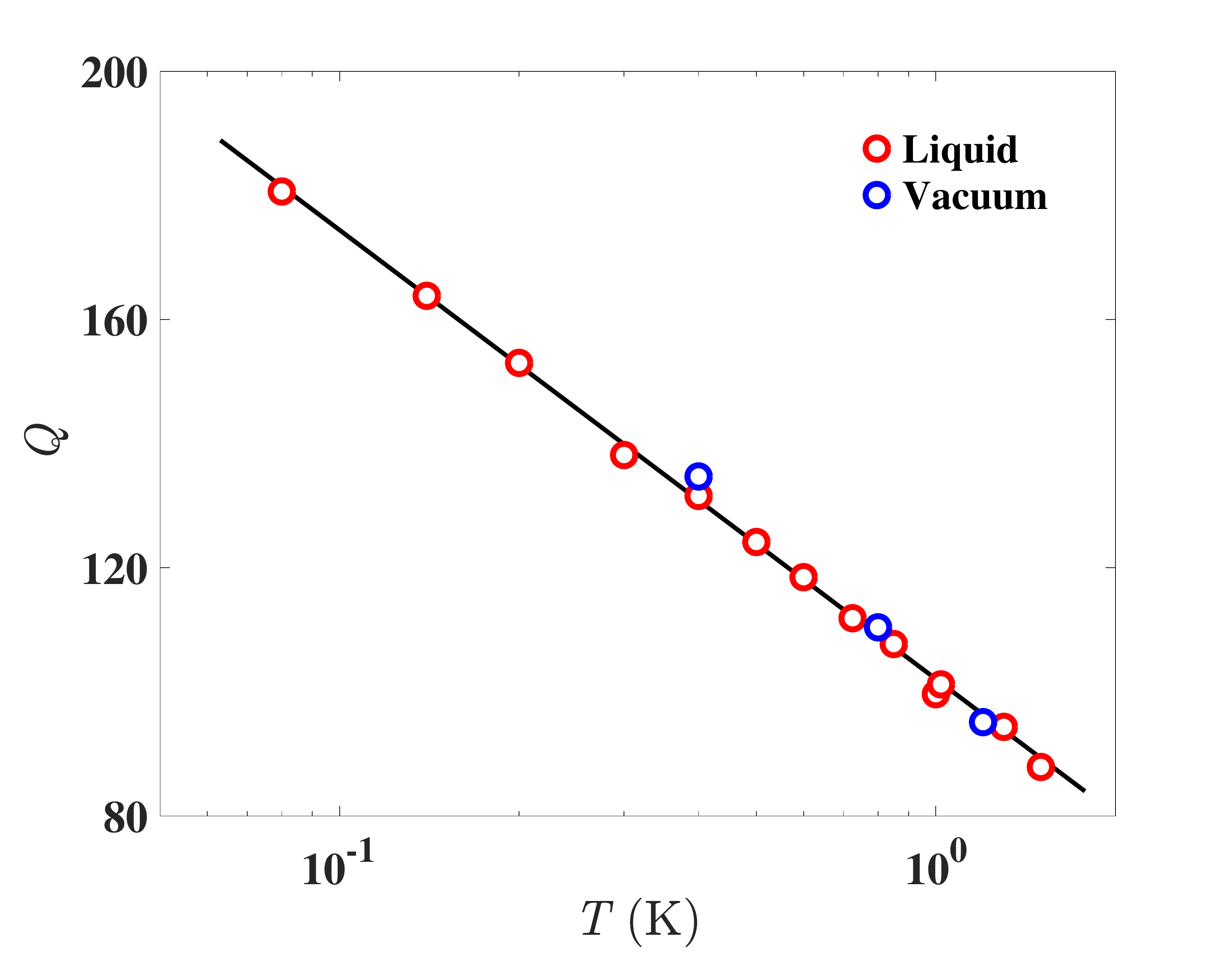}
\caption{(a) Center frequency $f_{\text{res}}$ of the surface mode
  around 22.5~MHz as a function of temperature. The solid line
  indicates $\log T$ dependence. (b) Quality factor of the surface
  mode as a function of temperature. The linear trend in the data
  indicates $\log (T)$ dependence. The sample chamber was filled with
  liquid at $p=0.5\,$bar. Blue circles denote reference data measured
  in vacuum.}
\label{fig:fT}
\end{figure}

Fig.~\ref{fig:fT}a displays the center frequency of the resonance of
Fig.~\ref{fig:resoShape2} as a function of temperature $T$; besides
data in superfluid at pressure $p=0.5\,$bar, there are also coinciding
reference data obtained in vacuum. The center frequency was extracted
using fits of Eq.~\ref{fitfunc} to the measured line shapes.  Clearly,
a logarithmic variation of the resonance frequency with temperature is
observed below 700~mK. Above 700~mK, the resonant frequency starts to
decrease with $T$, in a similar fashion as observed in many glassy
substances~\cite{Piche1974}.

The independence of the resonance frequency on the superfluid
environment indicates that the mode is likely a surface shear mode (a
generalized Love wave~\cite{McHale2002}) which couples to suspended
graphene via a change in the strain of the device. At low
temperatures, the fraction of normal fluid is small~\cite{Tilley1990},
and no apparent viscous coupling from superfluid is observed on the
shear mode. In the absence of any enhancement in the effective mass,
the mode frequency remains unchanged in superfluid $^4$He environment.

Fig.~\ref{fig:fT}b displays the quality factor $Q$ (see
Eq.~\ref{fitfunc}) of the surface mode at 22.5~MHz as a function of
temperature; the data points have been extracted from the same fits as
the center frequencies in Fig.~\ref{fig:fT}a. Our data indicate that
the dissipation of this mode is the same both in vacuum and in
superfluid $^4$He. It also means that the surface wave does not emit
sound waves in superfluid helium: because the sound velocity in liquid
helium is very small, about 200~m/s~\cite{Tilley1990}, there is a
strong acoustic mismatch against emitting sound from the surface wave
into the superfluid. Thus, the same $\log T$ dependent behavior of $Q$
is observed in helium liquid as in vacuum, and the quality factor is
found to change from $Q=90$ at $T=1000\,$mK up to $Q=170$ at $T=100\,$mK.

Similar results were also obtained with measurements in vacuum on
suspended monolayer samples and a non-suspended control sample. For
two suspended monolayer samples with the same $150 \times
150\,\upmu$m$^2$ bonding pad size as in the trilayer sample, broad
double peak resonances with quality factors $Q \sim 100$ were found at
21.0~MHz and 22.1~MHz, respectively. A similar resonance at
$f_{\text{res}}=$~21.2 MHz with $Q\approx 80$ was found in the
LOR-supported control sample. However, significantly higher RF input
power, $\sim$40~dB more $w.r.t$ the trilayer measurements, and
$\sim$20~dB more $w.r.t$ suspended monolayer case, was required to
produce the same mixing current in the LOR-supported sample.

\section{Discussion}\label{discussion}

Properties of glassy materials at low temperatures have been
investigated extensively in the past using surface waves~\cite{Piche1974,Hunklinger1980}: Strong logarithmic temperature
dependence of sound velocity has been obsereved in vitreous silica
glass, quite contrary to the behavior observed in fully crystalline
quartz. The observed $\log T$ variation was interpreted by Piche et
al.~\cite{Piche1974} as characteristics for amorphous materials due to
the existence of tunnelling defects
\cite{Anderson1972,Phillips1972}. Besides amorphous silicon oxide,
surface waves have been observed in amorphous germanium at
temperatures $0.1 - 10\,$K~\cite{Duquesne1983}, and a logarithmic
temperature correction in the sound velocity at 200 MHz was observed
below 1 K. Since our measured $f_{\rm res}(T)$ indicates similar $\log T$
dependent velocity variation as in
Refs.~\onlinecite{Piche1974,Hunklinger1980,Duquesne1983}, we associate
our findings with two level state systems in the amorphous resist layer,
which influence the acoustic properties across the whole sample,
including the effective speed for surface shear waves.  Note that, for
crystalline solids, $v_{\rm sh}(T)$ would be a decreasing function of
temperature as the bulk modulus is reduced with increased
thermally-induced lattice motion.

Owing to their special properties, polymers have given rise to great
interest as acoustic guiding layers for localized Love
waves~\cite{Gizeli1992a,Du1996,Harding1997,Roach2007}. To our
knowledge, however, Love modes with polymers have not yet been
investigated at cryogenic temperatures, although glassy properties of
PMMA at MHz frequencies have been addressed by Federle and Hunklinger
at $T = 0.06 - 10\,$K~\cite{Federle1982}. Our work is able to cast
light on the properties of generalized Love modes at low $T$ using
graphene-based detection, and thereby we can investigate glassy
properties of polymers in a new fashion.

 The coupling between surface phonons and the TLS systems contains two
 major contributions: 1) the resonant coupling and 2) relaxation
 processes \cite{Hunklinger1998}. When $\hbar \omega < k_{\rm B} T$, the
 resonant absorption can be neglected as both of the two level states
 are nearly equally populated. The resonance processes, nevertheless,
 contribute to elastic properties at low temperatures. These processes
 lead to a velocity correction given by (see
 e.g. Ref.~\onlinecite{Tate1979}):
\begin{equation}\label{deltav}
  \frac{\Delta  v}{ v}=\frac{n_0 M_{\rm s}^2}{\rho  v^2}K \log\left(\frac{T}{T_0}\right)= C \log\left(\frac{T}{T_0}\right),
\end{equation}
where $M_{\rm s}$ signifies the deformation potential, $n_0$ is
density of the tunnelling defects, $\rho$ is the mass density, $v$ is
the velocity of the sound wave, $K$ is a factor on the order of unity,
and $T_0$ denotes an arbitrary reference temperature. The change in
$\frac{\Delta v}{ v}$ will lead to a corresponding change in the
frequency of the resonance mode $\frac{\Delta
  f_{\text{res}}}{f_{\text{res}}} = \frac{\Delta v_{\rm sh}}{ v_{\rm
    sh}}$ plotted in Fig. \ref{fig:fT}a.  Note that the coupling
parameter $M_{\rm s}$ and its scaled value $C$ have to be considered
as averaged quantities because of the intrinsic non-uniformity of the
glassy resist materials in question~\cite{Federle1982}.

By fitting Eq.~\ref{deltav} to the data in Fig.~\ref{fig:fT}a, we
obtain $C=4 \times 10^{-3}$. The properties of PMMA resist (density
$\rho_{\rm PMMA} = 1.18\,$g/cm$^3$, $ v_{\rm PMMA} = 1600\,$m/s at
4.2~K) down to 60~mK have been investigated in
Ref.~\onlinecite{Federle1982}. A $\log T$ dependent variation of the
sound speed was found, which amounted to approximately to $\Delta v/v
\simeq 5 \times 10^{-4}$ over the range $60 - 500\,$mK. This is eight
times smaller than our result on a similar polymer LOR 3A (density
$\rho_{LOR} = 0.98\,$g/cm$^3$). Since $C \propto K(n_0/\rho) (M_{\rm s}/
v)^2$, this would suggest that both the deformation potential and the
density of two level states have to be larger in LOR 3A than in PMMA
($\rho$ and $ v$ are taken as approximately equal for LOR 3A and PMMA)
\footnote {Because of the high quality of our SiO$_2$ layer and that
  the thickness of SiO$_2$ is clearly less than that of LOR 3A, we
  neglect the influence of the SiO$_2$ layer in our analysis, although
  SiO$_2$ fulfills the speed condition for guiding a shear wave on top
  of silicon \cite{Du1998}}. Furthermore, according to
Ref. \onlinecite{Venkatesan2010}, $C \simeq 10^{-5}$ for gold beam
resonators and, thus the dissipation due to TLS systems in gold can be
neglected in our work.

At high temperatures, relaxation processes dominate over the resonant
ones, which leads to opposite sign for the prefactor $C$ in
Eq.~\ref{deltav}~\cite{Hunklinger1998}. Thus, there will be a maximum
in $\frac{\Delta f_{\text{res}}}{f_{\text{res}}}$ as a function of
$T$, which is observed in Fig. \ref{fig:fT}a. Furthermore, at
temperatures above the maximum of $\frac{\Delta
  f_{\text{res}}}{f_{\text{res}}}$, acoustic losses are mostly
dominated by processes having $\omega \tau \sim 1$, which would lead
to $Q^{-1}=\frac{\pi}{2}C$. This formula yields to $Q \simeq 160$
around 1 K which is close to the measured value $Q = 90$ obtained from
Fig.~\ref{fig:fT}b.

Figure~\ref{fig:fT}b displays a logarithmic decrease
  of the apparent quality factor $Q$ with increasing temperature.
  This is in contrast to the $\omega T^{-3}$ behavior governing the
  accoustic attenuation if only TLS are accounted
  for~\cite{Hunklinger1998}.  Here, other mechanisms dominate the
  observed behavior, the most obvious being the specific frequency
  depence of the dynamic compliance of the source
  configuration. Taking the pads into account, we have for a
  horizontally vibrated pad in the regime where $\omega_{\rm
    res}\gtrsim v_{\rm sh}/a$ (as is the case here), that the
  imaginary part of the dynamic compliance scales as $\omega_{\rm
    res}G$~\cite{Westman1972}. This leads to, consistent with observation, $\Delta Q\propto -\Delta
  v_{\rm sh}/v_{\rm sh} \propto -\log T$. Thus, the qualitative the
  origin of the apparent dissipative behavior in Fig.~\ref{fig:fT}b
  lies in TLS-induced velocity changes in the LOR 3A layer.

In addition, one may argue that the both the velocity
  and the prefactor $C$ in Eq.~\ref{deltav} differs slightly under
  the gold layer compared with the LOR 3A region without gold. This
  leads to a temperature dependent velocity difference causing an
  acoustic impedance mismatch, guiding the shear wave towards the
  detector. The larger $C$ is consistent with additional disorder
  generation caused by heating or stress, i.e., that evaporation of
  Cr/Au enhances the amount of TLS systems in the
  resist. Consequently, as the speed difference $\Delta v = v_{{\rm
      sh}_0}-v_{\rm sh}$ decreases logarithmically with increasing $T$
  the impedance matching is improved and leading to effective losses
  from regions below gold to be proportional to $const-\frac{\Delta
    v_{\rm sh}}{v_{\rm sh}}$.  This leads to a $\log T$ dependent
  enhancement in the loss of energy in the gold covered region, and
  the graphene detector indicates an effective $Q \propto -\log T$.

Besides shear modes owing to the polymer layer, we have considered the
possibility of having a surface acoustic wave propagating in graphene
\cite{Thalmeier2010}. However, the weakness of the gate dependence of
the resonance frequency rules out this scenario. On the other hand,
according to the review of Ref.~\onlinecite{Pohl2002}, the coupling
between TLS systems and surface waves can be large in polymers, which
supports our view that the obtained surface wave results describe
intrinsic TLS-induced properties of generalized Love waves in the
resist/substrate system.

In summary, without resorting to typical piezoelectric generation and
detection of surface acoustic waves, we have successfully investigated
propagating surface shear modes, generalized Love waves on
resist/silicon substrate system loaded by gold. These modes at
ultrasonic frequencies could be excited resonantly by the high
electric field below the gold bonding pads which acted as phonon
cavity resonators. For detection, we employed graphene as a suspended
(and non-suspended) strain gauge in a geometry which facilitated
separation between graphene, gold and surface shear modes. Proof of
the nature of the shear mode in the resist/substrate setting was
inferred from the insensitivity of the mode frequency and the losses
upon immersion of the device into superfluid helium. Our data on the
temperature dependence of the propagation properties of this shear
mode indicate strong coupling between two level systems and the
surface wave. The coupling can be interpreted in terms of the regular
tunneling TLS model in the presence of strong disorder.

\section*{Acknowledgements}
We thank F. Massel for fruitful discussions. This work has been supported by the Academy of Finland (projects no. 314448 and 310086) and by ERC (grant no. 670743). This research was funded by European Union's Horizon 2020 Research and Innovation Programme under Grant Agreement No. 824109. A.L. and J.-P.K. are grateful to Vais{\"a}l{\"a} Foundation of the Finnish Academy of Science and Letters for a scholarship. A.I. acknowledges the hospitality of Aalto University. This research project made use of the Aalto University OtaNano/LTL infrastructure which is part of European Microkelvin Platform.

\bibliography{Collection}

\end{document}